\documentclass[a4paper,12pt]{article}

\begin{document}

\noindent{\it\small Astronomical and Astrophysical Transactions}

\noindent{\small Vol.25, No.2, Month 2006}

\vskip 2cm

\begin{center}

%title
\textbf{\large INVESTIGATION OF NEGATIVE K-EFFECT USING THE OSACA CATALOGUE}\\

\bigskip

%authors
\large{V.~V.~Bobylev*,  A.~T.~Bajkova and G.~A.~Gontcharov}\\

\bigskip

\normalsize

%address
Main (Pulkovo) Astronomical Observatory of the Russian Academy of
Sciences, Pulkovskoye
Chaussee 65/1, St-Petersburg 196140,  Russia\\

\bigskip

%e-mail
*Corresponding author. Email: vbobylev@gao.spb.ru\\

\vskip 1cm

{\it\small (Received 29 June 2006)}

\vskip 1 cm

\end{center}

\small
%abstract
\noindent We created and are constantly updating the Orion Spiral
Arm CAtalogue (OSACA) for stars with known coordinates,
parallaxes, proper motions and radial velocities. It is shown that
there is an effect of contraction appearing in motion of giants of
A0--A5 spectral classes which attains a value of $K=-13\pm2$~km
s$^{-1}$ kpc$^{-1}$. We try to link this effect to the periodic
structure of residual velocity field of stars in the solar
neighborhood, which is caused by spiral density waves.\\

\bigskip

%keywords

\bigskip

\normalsize
%main text
\noindent In the frame of Ogorodnikov-Milne model the presence of
negative $K$-effect  means  that the  star system being considered
is found in the state of contraction. The negative $K$-effect of
value $-(1\div7)$~km s$^{-1}$ kpc$^{-1}$ was found in the motion
of OB stars by  Fern\'andez {\it et al.} [1], Bobylev [2], Rybka
[3]. At present, the nature of this phenomenon is not completely
clear. There are different hypotheses:  this effect  is connected
with specific measurements of star radial velocities [4], or with
 the influence of the bar in the Galactic centre [4], or with the influence of
the spiral structure [5].

The goal of this work is to establish the connection of the
negative $K$-effect with the periodicity of star's velocity field
caused by the spiral structure.

The database used here represents the  radial velocities of stars
collected from more than 1400 bibliographical sources. They are
reduced to a single system of radial velocities of 854 standard
stars from the list formed by us. This allowed us to calculate
weighted-mean radial velocities with a median accuracy of
$\pm1$~km s$^{-1}$ for more than 25~000 Hipparcos stars  located
in the Orion Arm region. Together with other characteristics,
these values are represented in the form of a constantly updated
database and the Orion Spiral Arm Catalogue  (OSACA) [6,7]).

We used 1269 stars of A0--A5 spectral classes with B$-$V$\leq0.2$.
Stars of only I, II and III luminosity classes are included into
this sample. However, the basic part of the sample consists of
OSACA stars, without any notification of the luminosity class. All
selected stars belong to the distance interval $r=0.1-0.6$ kpc.
The velocities of the stars were corrected for the common Galactic
rotation with Oort constants $A=13.7$~km s$^{-1}$ kpc$^{-1}$ and
$B=-12.9$~km s$^{-1}$ kpc$^{-1}$  [2].

As a result, the following values of solar peculiar velocity were
found:
 $u_\odot=10.2\pm0.4$~km s$^{-1}$,
 $v_\odot=10.9\pm0.4$~km s$^{-1}$,
 $w_\odot= 6.6\pm0.4$~km s$^{-1}$.

 The roots of the deformation tensor are
 ($\lambda_1,\lambda_2,\lambda_3)=(3.5,-29.3,-3.5)$~km s$^{-1}$ kpc$^{-1}$, and their errors
 are approximately 3~km s$^{-1}$ kpc$^{-1}$.
Let one of the roots ($\lambda_3$) be zero, and consider only the
$x-y$ plane. We take into account that
 $(\partial V_R/\partial R)_{R_\circ}=\lambda_1$,
                   $(V_R/R)_{R_\circ}=\lambda_2$, where $R$ is
the distance from the kinematic centre, which is unknown. Using
the relations
 $K+C=4\pm3$~km s$^{-1}$ kpc$^{-1}$, $K-C=-29\pm3$~km s$^{-1}$
 kpc$^{-1}$,we can evaluate the typical residual velocity of stars as
 $(K-C)\cdot {\overline r}=-5.4\pm0.6$~km s$^{-1}$.
Contraction takes place along the axis in the direction
$12-192^\circ$.

To take into account the spiral structure, we have used the
following parameters found in work by Mel$'$nik {\it et al.} [8]:
 $f_R=6.6$~km s$^{-1}$,
 $f_\theta=1.8$~km s$^{-1}$,
 $\phi_R=38^\circ$,
 $\phi_\theta=-33^\circ$,
 $\lambda=2$~kpc
 at
 $R_\circ=7.1$~kpc and
 $i=6^\circ$.

For these parameters we found that
 $u_\odot=  4.2\pm0.5$~km s$^{-1}$,
 $v_\odot= 11.0\pm0.5$~km s$^{-1}$,
 $w_\odot=  6.7\pm0.5$~km s$^{-1}$.
The roots of the deformation tensor are
($\lambda_1,\lambda_2,\lambda_3$)=($2.0,-33.5,-3.9)$~km s$^{-1}$
kpc$^{-1}$, and  $K=-15.9\pm2.2$~km s$^{-1}$ kpc$^{-1}$.

For the second case we have used the spiral structure parameters
found in the work by Popova and Loktin [9]:
 $f_R=-3.97$~km s$^{-1}$ and
 $f_\theta=13.27$~km s$^{-1}$,
 at
 $R_\circ=8.3$~kpc and
 $i=21.5^\circ$.
 Our results are as follows:
 $u_\odot=  8.6\pm0.5$~km s$^{-1}$,
 $v_\odot= -1.3\pm0.5$~km s$^{-1}$ and
 $w_\odot=  6.7\pm0.5$~km s$^{-1}$.
  The roots of the deformation tensor are ($\lambda_1,\lambda_2,\lambda_3$)
=($5.4,-29.8,-3.5)$~km s$^{-1}$ kpc$^{-1}$, and $K=-12.1\pm2.2$~km
s$^{-1}$ kpc$^{-1}$. A comparison of these two results shows that
the negative $K$-effect exists in the both the cases considered.

\bigskip

\noindent{\bf Acknowledgement}

\bigskip

\noindent{This work is supported by the Russian Fund for Basic
Research (grant No~05--02--17047).}


\bigskip

\begin{thebibliography}{8}

\bigskip

\bibitem{1}

D. Fern\'andez, F. Figueras, and J. Torra, Astron. Astrophys. {\bf
372}\rm\, 833 (2001).

\bibitem{2}

V. V.~Bobylev, Pis$'$ma Astron. Zh. {\bf 30}\rm\, 185 (2004).

\bibitem{3}

S. P. Rybka, Kinem. Phys. Neb. Tel. {\bf 20}\rm\, 437 (2004).

\bibitem{4}

F. Pont, M. Mayor, and G. Burki, Astron. Astrophys. {\bf 285}\rm\,
415 (1994).

\bibitem{5}

M. R. Metzger, J. A. R. Caldwell, and P. L. Schechter, Astron. J.
{\bf 115}\rm\, 635 (1998).

\bibitem{6}

Orion Spiral Arm Catalogue, http://www.geocities.com/orionspiral/
(2006).

\bibitem{7}

V. V. Bobylev,  G. A. Goncharov, and  A. T. Bajkova, Astron. Zh.
{\bf 83}\rm\, 821 (2006).

\bibitem{8}

A. M. Mel$'$nik, A. K. Dambis, and A. S. Rastorguev, Pis$'$ma
Astron. Zh. {\bf 27}\rm\, 611 (2001).

\bibitem{9}

M. E. Popova, A. V. Loktin, Pis$'$ma Astron. Zh. {\bf 31}\rm\, 743
(2005).

\end{thebibliography}
\end{document}